\documentclass[aps,prl,showpacs,twocolumn,amsmath,amssymb,superscriptaddress,footinbib]{revtex4}

\usepackage[english]{babel}
\usepackage{latexsym}
\usepackage{graphics}
\usepackage{subfigure}
\usepackage{epsfig}
\usepackage{color}

\def\be{\begin{equation}}
\def\ee{\end{equation}}
\def\bea{\begin{eqnarray}}
\def\eea{\end{eqnarray}}
\def\bi{\begin{itemize}}
\def\ei{\end{itemize}}
\def\bin{\begin{enumerate}}
\def\ein{\end{enumerate}}

\def\la{\langle}
\def\ra{\rangle}

\newcommand{\vect}[1]{\mathbf{#1}}

\begin{document}

\title{Disorder-Induced Phase Control in Superfluid Fermi-Bose Mixtures}

\author{A.~Niederberger} \affiliation{ICFO-Institut de Ci\`encies Fot\`oniques, Parc Mediterrani de la Tecnologia, E-08860 Castelldefels (Barcelona), Spain}
\author{J.~Wehr} \affiliation{ICFO-Institut de Ci\`encies Fot\`oniques, Parc Mediterrani de la Tecnologia, E-08860 Castelldefels (Barcelona), Spain} \affiliation{Department of Mathematics, The University of Arizona, Tucson, AZ 85721-0089, USA}
\author{M.~Lewenstein} \affiliation{ICFO-Institut de Ci\`encies Fot\`oniques, Parc Mediterrani de la Tecnologia, E-08860 Castelldefels (Barcelona), Spain} \affiliation{ICREA - Instituci\`o Catalana de Ricerca i Estudis Avan{\c c}ats, E-08010 Barcelona, Spain}
\author{K.~Sacha} \affiliation{ICFO-Institut de Ci\`encies Fot\`oniques, Parc Mediterrani de la Tecnologia, E-08860 Castelldefels (Barcelona), Spain} \affiliation{Marian Smoluchowski Institute of Physics and Mark Kac Complex Systems Research Centre, Jagiellonian University, Reymonta 4, PL-30059 Krak\'ow, Poland}

\date{\today}

\begin{abstract}
We consider a mixture of a superfluid Fermi gas of ultracold atoms and a
Bose-Einstein condensate of molecules possessing a continuous $U(1)$ (relative phase) symmetry.  We study the effects  that a spatially random photo-associative-dissociative symmetry breaking coupling of  the systems. Such coupling allows to control the relative phase between a superfluid order parameter of the Fermi system and the condensate wavefunction of 
molecules for temperatures below the BCS critical temperature. 
The presented mechanism of phase control belongs to the general class of
disorder-induced order phenomena that rely on breaking of continuous symmetry.

\end{abstract}


\maketitle

Of all the studies of the physics of ultracold atoms, fermionic systems and in particular Fermi superfluids have recently received a lot of attention. The most famous example in this context is the BEC-BCS crossover. In this example, fermions are transformed between the superconducting state in which they form Bardeen-Cooper-Schriefer (BCS) pairs and the Bose-Einstein Condensate (BEC) made of diatomic molecules of the same fermions (for reviews see \cite{varrena,bloch,pita}). Another field of particularly high activity has recently been the study of disorder in ultracold gases. Since its proposal \cite{damski}, several experimental and theoretical groups have focused on this subject \cite{Advan,inguscio}. Very recently, these studies culminated in the seminal experimental observation of Anderson localization of matter waves in a BEC with a disordered potential \cite{nature1,nature2} and in signatures of the Bose glass state of an ultracold gas in an optical lattice \cite{fallani}.

We have proposed an additional type of disorder effects that can be realized with ultracold atoms, called {\it disorder induced order} \cite{our}. Generally speaking, the effect can be observed in systems possessing a continuous symmetry which, in accordance with the Mermin-Wagner-Hohenberg theorem, prevents long range order in low dimensions \cite{mwh}. Comparing models without and with continuous symmetries, we therefore observe that the latter have lower critical temperatures in any dimension. Adding disorder that breaks the continuous symmetry leads to competing mechanisms in such systems: naturally, disorder acts against ordering whereas symmetry breaking increases the tendency to order. We have studied this effect for a classical $XY$ model in a random $X$ oriented magnetic field and have shown rigorously the appearance of spontaneous magnetization in the $Y$ direction at $T=0$. Also, we have presented strong evidence for the survival of magnetization at $T>0$ in the limit of small disorder. These ideas were then applied to quantum systems: a two component Bose gas in an optical lattice \cite{our} and a two component BEC with random Raman coupling \cite{prl}. It is worth noticing that similar effects can be found in other areas of physics (see for instance \cite{dupa-prl}, for classical mean field modelsÊsee \cite{sens}). Also, the effect is robust enough to appear not only in random, but also in oscillating symmetry breaking perturbations.

The main goal of the present letter is to introduce these new ideas to the field of superfluid Fermi
gases that posses a phase $U(1)$ symmetry.  We show that a Fermi superfluid coupled to a molecular BEC via a random, symmetry breaking photo-associating-dissociating coupling  undergoes relative phase ordering, so that the phase of the order parameter can be efficiently controlled by the phases of the coupling. Within a BCS-like theory \cite{fetter}, we show rigorously that for small disorder the effect occurs practically for all temperatures below the superfluid transition temperature. 

We consider a mixture of fermions in two different internal
states interacting via attractive zero-range potential in a volume $V$ 
in 3D with a Hamiltonian
\be
H_F=\int d\vect r \left[ \hat\psi_j^\dagger  \left(-\frac{\hbar^2}{2m}\nabla^2-\mu
\right) \hat\psi_j - g\hat\psi_1^\dagger\hat\psi_2^\dagger\hat\psi_2\hat\psi_1 
\right],
\label{hf}
\ee  
where the chemical potential $\mu$ fixes the average density $n$, 
$g=4\pi\hbar^2|a|/m$ (with s-wave scattering length $a$) determines strength of
the interactions and the $\hat\psi_j$ stand for fermionic field operators. 
Moreover, we assume the presence of a Bose-Einstein condensate
(BEC) of molecular dimers consisting of the two fermionic species and a (weak) coupling  that transforms them into
fermion pairs  and vice versa.
Experimentally, we can sweep a mixture of fermions with two different
internal states over a Feshbach resonance, which leaves us with a BEC of diatomic
molecules and unbound fermions. Then, we approach a second Feshbach resonance
which turns the previously unbound fermions into BCS pairs
without affecting the molecular BEC. The coupling between the molecules and the fermions
can be realized through photoassociation and photodissociation, respectively. 
Taking the limit of a large BEC, we do 
not need to consider the dynamics of the BEC because the 
effect of the weak coupling with the fermions on BEC is negligible. Following these 
considerations, 
the Hamiltonian (\ref{hf}) has to be
supplemented by one term only: the coupling between the fermions and the
BEC which we approximate as 
$\int d\vect r  \left[\gamma^*(\vect r)\hat\psi_d^\dagger\hat\psi_1\hat\psi_2+
\gamma(\vect r)\hat\psi_2^\dagger\hat\psi_1^\dagger\hat\psi_d\right]\approx
\sqrt{n_d}\int d\vect r  \left[\gamma^*(\vect r)\hat\psi_1\hat\psi_2+
\gamma(\vect r)\hat\psi_2^\dagger\hat\psi_1^\dagger\right]$, where the bosonic
filed operator $\hat\psi_d$ for molecules is substituted by a real-valued condensate 
wavefunction which for
a homogeneous case considered here is squared root of the density of dimers 
$\sqrt{n_d}$. The full Hamiltonian therefore reads
\be
H=H_F+\int d\vect r  \left[\Gamma^*(\vect r)\hat\psi_1\hat\psi_2+
\Gamma(\vect r)\hat\psi_2^\dagger\hat\psi_1^\dagger\right],
\label{h}
\ee
with $\Gamma(\vect r)=\tilde\Gamma(\vect r)e^{-i\varphi_\Gamma}
=\sqrt{n_d}\gamma(\vect r)$. We assume that the transfer
process is realized such that $\tilde\Gamma(\vect r)$ is real, 
changes randomly in space and is constant in time, $\int d\vect r \tilde\Gamma(\vect r)=0$ and
$\varphi_\Gamma$ is a real number. We will show that for $\varphi_\Gamma=0$
the relative phase between the condensate wavefunction of molecules and the paring 
function of the superfluid fermions is fixed to $\pi/2$ (or $-\pi/2$). Then we show that
we can control the relative phase and fix it to any value by changing
a control parameter $\varphi_\Gamma$. 

For $\Gamma=0$ and in the weak coupling limit, i.e. $g\rightarrow 0$, 
we deal with a Fermi system which for $T$ less than the critical 
temperature $T_c=8e^\eta e^{-2}\pi^{-1} T_Fe^{-\pi/2k_F|a|}$ 
(where
$k_BT_F=\varepsilon_F=\hbar^2k_F^2/2m=\hbar^2(3\pi^2n)^{2/3}/2m$ and
$\eta=0.5772$) reveals a transition 
to a superfluid phase (BCS state) which is indicated by a non-vanishing pairing 
function (order parameter) $\Delta=g\la\hat\psi_2\hat\psi_1\ra$
\cite{randeria,fetter}. In the general
case (i.e. including non-zero $\Gamma$) the pairing function is given 
in terms of 
solutions of Bogoliubov-de~Gennes (BdG) equations, i.e. 
$\Delta=g\sum_n u_nv_n^*[1-2f(E_n)]$, where
\be
\left[
\begin{array}{cc}
 -\frac{\hbar^2\nabla^2}{2m}-\mu+W & \Delta+\Gamma \\
\Delta^*+\Gamma^*  & \frac{\hbar^2\nabla^2}{2m}
+\mu-W
\end{array}
\right]
\left[
\begin{array}{c}
u_n \\ v_n
\end{array}
\right]
=
E_n
\left[
\begin{array}{c}
u_n \\ v_n
\end{array}
\right],
\label{bdg}
\ee
with a Hartree-Fock term 
$W=-g\la\hat\psi_1^\dagger\hat\psi_1\ra=-g\la\hat\psi_2^\dagger\hat\psi_2\ra=-g\sum_n \left(|u_n|^2f(E_n)+|v_n|^2[1-f(E_n)]\right)$ and
$f(E_n)=\left(e^{E_n/k_BT}+1\right)^{-1}$ \cite{fetter}.

If the transfer process is absent (i.e. $\Gamma=0$) the system (\ref{h}) is 
invariant under global gauge transformation, i.e. 
$\hat\psi_j\rightarrow e^{i\varphi/2}\hat\psi_j$, which implies
that if $\{u_n,v_n\}$ are solutions of the BdG equations for $\Delta$, then 
$\{e^{i\varphi/2}u_n,e^{-i\varphi/2}v_n\}$ are the solutions corresponding to 
$e^{i\varphi}\Delta$.
This continuous symmetry is broken when the
transfer process is turned on, as can be seen through (\ref{h}). Then the phase of the pairing
function becomes relevant because it is the relative phase with respect to 
the (real-valued) condensate wavefunction of dimers. 

We begin with an analysis of the $\varphi_\Gamma=0$ case, i.e. 
$\Gamma=\tilde\Gamma$ is real.
Let us assume that for $\Gamma=0$ and for
some temperature $T$ we have a non-zero pairing function which is chosen to be 
real and positive, $\Delta_0>0$. When we turn on $\Gamma$ but 
$|\Gamma(\vect r)|\ll \Delta_0$ one may expect that it results in a new pairing 
function where $\Delta(\vect r)\approx \Delta_0e^{i\varphi(\vect r)}$. That is, 
any non-zero $\Gamma$ has a dramatic effect on the phase because
without the transfer process the system is degenerated with respect to the
choice of $\varphi$. On the other hand an infinitesimally small $\Gamma$ is 
not able to change $|\Delta(\vect r)|$ because this would cost energy. Moreover, we 
may expect that $\varphi(\vect r)$ oscillates around some average value
$\varphi_0$ with small amplitude because we assume that
$\Gamma(\vect r)$ fluctuates around
zero with infinitesimally small variance.
Under these assumptions we can observe that $|\varphi_0|=\pi/2$.
Indeed, let us neglect the Hartree-Fock term $W$ (which is not essential for 
Fermi superfluidity) and calculate the difference of the thermodynamic
potentials for superfluid and normal phase 
\bea
\Omega_s-\Omega_0&=& \int_0^1\frac{d\lambda}{\lambda}
\la \lambda H_1\ra_\lambda
\cr
&\approx&
 -\int_0^{g}\frac{dg'}{{g'}^2}\int d\vect r 
\left[|\Delta|^2 
+\frac{2g'}{g}
\tilde\Gamma|\Delta|\cos\varphi\right],
\label{omega}
\eea
where $H_1=\int d\vect r 
\left[\tilde\Gamma(\hat\psi_1\hat\psi_2+
\hat\psi_2^\dagger\hat\psi_1^\dagger)
-g\hat\psi_1^\dagger\hat\psi_2^\dagger\hat\psi_2\hat\psi_1\right]$ 
\cite{fetter}.
According to the assumptions, for $g'$ close to $g$, $|\Delta(\vect r)|$ is constant and  
$\cos\varphi(\vect r)\approx \cos\varphi_0
-\sin\varphi_0\delta\varphi(\vect r)$. Then
\bea
-\int d\vect r \left[|\Delta|^2+\frac{2g'}{g}
\tilde\Gamma|\Delta|\cos\varphi\right] 
\approx 
\cr
-|\Delta|^2V
+\sin\varphi_0\frac{2g'|\Delta|}{g}
\int d\vect r \tilde\Gamma\delta\varphi,
\eea
and for $\int d\vect r \tilde\Gamma\delta\varphi<0$ the thermodynamic potential 
is minimized when $\varphi_0=\pi/2$. With the 
transformation $\delta\varphi\rightarrow -\delta\varphi$ and 
$\varphi_0\rightarrow-\pi/2$ we obtain another solution what 
reflects a symmetry of the system. That is, for a real $\Gamma$, 
if $\Delta(\vect r)$ corresponds to solutions of (\ref{bdg})  
then solutions of complex conjugate 
BdG equations results in a new pairing function equal $\Delta^*(\vect r)$.
In experiments the sign of $\varphi_0$ will depend on the preparation  
and is determined by spontaneous breaking of the
$\varphi\rightarrow-\varphi$ symmetry.

Having determined $\varphi_0$ we would like to estimate fluctuations of the 
phase of the pairing function $\delta\varphi(\vect r)$. To this end let us
employ Ginzburg-Landau (GL) approach \cite{fetter}. 
Adapting the Gorkov's derivation of the GL equation \cite{gorkov,fetter,baranov} 
(with the standard regularization of the bare interaction $g$ for the case of cold atomic gases) to our problem one obtains
\bea
\nabla^2\Delta=-\nabla^2\tilde\Gamma-\frac{48\pi^2}{7\zeta(3)l_c^2}
\left(\frac{2\pi^2\hbar^2}{mk_Fg}+\frac{T_c-T}{T_c}\right)\tilde\Gamma \cr
-\frac{48\pi^2}{7\zeta(3)l_c^2}\frac{T_c-T}{T_c}\Delta
+\frac{6m^2}{\hbar^4k_F^2}|\Delta+\tilde\Gamma|^2(\Delta+\tilde\Gamma),
\label{gl}
\eea
where $l_c=\hbar^2 k_F/mk_BT_c$. Equation~(\ref{gl}) is
valid for $T_c-T\ll T_c$ and for $\tilde\Gamma(\vect r)$ 
that changes on a scale much larger than $l_c$ (e.g. for $k_F|a|=0.5$ and
$n\sim10^{14}$~cm$^{-3}$ we get $l_c\sim 4~\mu$m).
For $|\tilde\Gamma(\vect r)|$ much smaller than $|\Delta_0(T)|$, i.e. the pairing
function in the absence of the transfer process, we may introduce further
approximations that 
allow us to reduce Eq.~(\ref{gl}) to
\be
|\Delta_0|\nabla^2\delta\varphi(\vect r)=\nabla^2\tilde\Gamma(\vect r)+\frac{48\pi^2}{7\zeta(3)l_c^2}
\left(\frac{2\pi^2\hbar^2}{mk_Fg}+\frac{T_c-T}{T_c}\right)\tilde\Gamma(\vect r),
\label{gl1}
\ee
where $|\Delta|\approx|\Delta_0|$ and 
we have chosen $\varphi_0=\pi/2$ in the expansion $\varphi(\vect
r)\approx\varphi_0+\delta\varphi(\vect r)$.
The solution of (\ref{gl1}) reads
\be
\delta\varphi(\vect k)=\frac{\tilde\Gamma(\vect k)}{|\Delta_0|}
-\frac{48\pi^2}{7\zeta(3)l_c^2|\Delta_0|}
\left(\frac{2\pi^2\hbar^2}{mk_Fg}+\frac{T_c-T}{T_c}\right)\frac{\tilde\Gamma(\vect
k)}{|\vect k|^2},
\label{sgl}
\ee
in the Fourier space.

Now we switch to a general case of complex 
$\Gamma=\tilde\Gamma e^{-i\varphi_\Gamma}$. It is easy to
check that if $|\Delta|e^{i\varphi}$ corresponds to solutions of 
the BdG equations with $\varphi_\Gamma=0$ 
then $|\Delta|e^{i(\varphi-\varphi_\Gamma)}$ 
is related to the solutions for $\varphi_\Gamma\ne 0$. 
This implies that, if for $\varphi_\Gamma=0$ we are able to fix the relative phase
between the condensate wavefunction of molecules and the pairing function of the
superfluid fermions to $\pi/2$ (or $-\pi/2$), then changing $\varphi_\Gamma$
allows us to fix it to $\phi_0=\pi/2-\varphi_\Gamma$
(or $\phi_0=-\pi/2-\varphi_\Gamma$) and phase control emerges.

Assuming that the transfer process with small $|\Gamma(\vect r)|$ results in 
phase fluctuations of $\Delta(\vect r)$ only, we have shown that the
fluctuations occur around $\pi/2-\varphi_\Gamma$ (or $-\pi/2-\varphi_\Gamma$)
and they are given by Eq.~(\ref{sgl}). Now we would like to switch to
numerical solutions of the BdG equations (where, in contrast to the analytical
analysis, we do not neglect the Hartree-Fock term $W$) 
to demonstrate that indeed for $|\Gamma|\ll|\Delta_0|$ 
the fluctuations are small and the predicted phase control 
is possible. In 3D calculations we regularize the coupling constant $g$ in 
$\Delta=g\la\hat\psi_2\hat\psi_1\ra$, i.e.
$g\rightarrow g_{eff}$, according to
\be
\frac{1}{g_{eff}}=\frac{1}{g}-\frac{mk_F}{2\pi^2\hbar^2}\left(
\frac12\ln\frac{\sqrt{E_C}+\sqrt{\varepsilon_F}}{
\sqrt{E_C}-\sqrt{\varepsilon_F}}-
\sqrt{\frac{E_C}{\varepsilon_F}}\right),
\ee
where the logarithm term results from the sum over Bogoliubov modes
corresponding to energy above the cut-off $E_C$ performed in the spirit of 
the local density approximation, see \cite{LDA} for details.
For the simulations we choose $L_z=40k_F^{-1}$, 
$L_\perp=20k_F^{-1}$, $\mu=0.83\varepsilon_F$ and $k_F|a|=0.4$ which 
for $\Gamma=0$ and the cut-off $E_C=100\varepsilon_F$ 
leads to $\Delta_0(T=0)=0.036\varepsilon_F$
and $T_c=0.019T_F$. 
Using these parameters $l_c\sim 100k_F^{-1}$ is larger than the
system size and we are able to explore a regime beyond GL theory.
We assume real 
$\Gamma(\vect r)$ given by a pseudo-random function that changes along 
$z$ direction only, 
\be
\Gamma(\vect r)=
\frac{\Gamma_0}{2} \! \left[\sin\! \left( \frac{2 \pi}{L_z} (9 z + 8.8) \! \right)
+\sin\! \left(\frac{2 \pi}{L_z} (13 z + 3.6) \! \right) \! \right].
\label{g}
\ee 

\begin{figure}
\centering
\includegraphics*[width=8.5cm,clip]{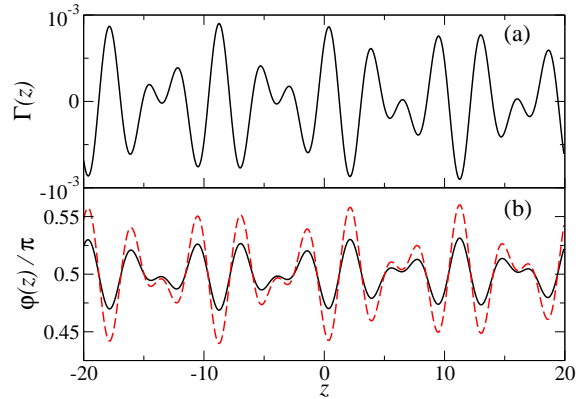}
\caption{
Panel (a) shows $\Gamma(z)$ given in Eq.~(\ref{g}) for 
$\Gamma_0=0.01|\Delta_0(0)|$. Panel~(b) presents the corresponding 
phase $\varphi(z)$ of the pairing function for $T=0$ (black solid curve) 
and $T=0.9T_c$ (red dashed curve).
}
\label{one}
\end{figure}

\begin{figure}
\centering
\includegraphics*[width=8.5cm,clip]{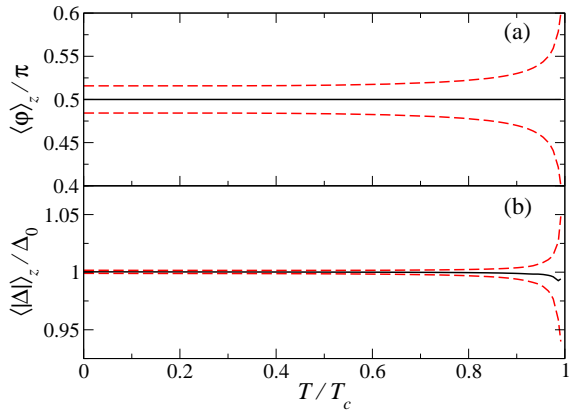}
\caption{
Panel~(a) shows average value of the phase of the pairing function 
$\la \varphi\ra_z$ versus temperature obtained for $\Gamma$ as in 
Fig.~\ref{one}. In panel~(b) we present the corresponding average value 
of the modulus of the pairing function $\la |\Delta|\ra_z$ divided by 
$\Delta_0(T)$, i.e. the pairing for the $\Gamma=0$ case. Solid black curves are
related to average values, dashed red curves to average values $\pm$
standard deviation. The figure shows simulations for temperatures up to 
$T=0.99T_c$ where $\Delta_0(T)=0.16\Delta_0(0)$.
}
\label{two}
\end{figure}

In Fig.~\ref{one} we show the phase of the pairing function $\varphi(z)$ 
in the case when
$\Gamma_0=0.01|\Delta_0(0)|$
and $\varphi_\Gamma=0$ for two different
temperatures, i.e. $T=0$ and $T=0.9T_c$. One can see 
that indeed the phase oscillates
around $\pi/2$ with a small amplitude (standard deviation of the order of
$10^{-2}$). The fluctuations of the absolute value of $\Delta(z)$
are negligible (standard deviations divided by average values are of the 
order $10^{-4}$). When $T$ approaches $T_c$ the average $|\Delta|$ decreases
and, at some $T$, becomes much smaller than $\Gamma_0$ and we enter another regime where 
the transfer term in the Hamiltonian (\ref{h}) starts dominating. 
For a very large $\Gamma_0$ we may expect that a real-valued $\Delta$, which fluctuates out
of the phase of $\Gamma(z)$, minimizes the thermodynamic potential. In
Fig.~\ref{two} we present average values and standard deviations for $\varphi$ 
and $|\Delta|$ versus temperature where one can observe an increase
of the fluctuations for $T\rightarrow T_c$.

In summary, we have shown how to control the relative phase $\varphi$ between the
wavefunction of a molecular condensate and the pairing function of a mixture
of fermions in the BCS state. It turns out that a certain class of weak couplings which transfer
pairs of fermions into molecules and vice versa, fix this relative
phase. The couplings need to vary  randomly, or in an oscillatory manner in space; they can be realized
by optical means, with a desired phase and amplitude, which in turn allows for efficient control of $\varphi$.  
In this letter we have considered the Fermi system in a weak coupling regime
but similar behavior is expected in the strong regime. In particular, translation of our results to the simplified resonant superfluidity theory (cf. \cite{holland})
is straightforward.
Our results hold also for $0<k_Fa\ll 1$,  where  
the pairing function becomes a condensate wavefunction of tightly bound pairs
and we deal with the control of relative phase between two Bose-Einstein condensates, analyzed in our earlier publication \cite{prl}. 
The problem considered here belongs to a general effect of disorder-induced 
order phenomena, that rely on continuous symmetry breaking.


We acknowledge support of the EU IP Programme `SCALA', the ESF Programmes "Qudedis", "Fermix", the Spanish MEC grants (FIS 2005-04627, Conslider Ingenio 2010 `QOIT'), 
J.W. was partially supported by the NSF grant DMS 0623941.
K.S. acknowledges Polish Government scientific funds (2008-2011) as 
a research project and by Marie Curie ToK project COCOS (MTKD-CT-2004-517186).
The research is partially conducted within LFPPI network.



\end{document}